\documentclass[sigconf,nonacm,natbib=true]{acmart}
\usepackage{subcaption}
\usepackage{graphicx}
\usepackage{makecell} 
\usepackage{colortbl}
\usepackage{xcolor}
\renewcommand\footnotetextcopyrightpermission[1]{}
\AtBeginDocument{%
  }

\begin{document}

\title{LLM-supported document separation for printed reviews from zbMATH Open}

\author{Ivan Pluzhnikov}
%\email{ivan.pluzhnikov@stud.uni-goettingen.de}
% \orcid{0000-0003-3219-026X}
\affiliation{
  \institution{George August University of Göttingen}
  \streetaddress{Wilhelmsplatz 1}
  \city{Göttingen}
  \country{Germany}
  \postcode{37073}
}

\author{Ankit Satpute}
\email{Ankit.Satpute@fiz-karlsruhe.de}
\orcid{0000-0003-3219-026X}
\affiliation{
  \institution{FIZ Karlsruhe Leibniz Institute for Information Infrastructure}
  \streetaddress{Franklinstrasse 11}
  \city{Berlin}
  \country{Germany}
  \postcode{10587}
}

\author{Moritz Schubotz}
\email{Moritz.Schubotz@fiz-karlsruhe.de}
\orcid{0000-0001-7141-4997}
\affiliation{
  \institution{FIZ Karlsruhe Leibniz Institute for Information Infrastructure}
\streetaddress{Franklinstrasse 11}
  \city{Berlin}
  \country{Germany}
  \postcode{10587}
}

\author{Olaf Teschke}
\email{Olaf.Teschke@fiz-karlsruhe.de}
\orcid{0009-0003-4089-9647}
\affiliation{
  \institution{FIZ Karlsruhe Leibniz Institute for Information Infrastructure}
  \streetaddress{Franklinstrasse 11}
  \city{Berlin}
  \country{Germany}
  \postcode{10587}
}

\author{Bela Gipp}
\email{gipp@uni-goettingen.de}
\orcid{0000-0001-6522-3019}
\affiliation{
  \institution{George August University of Göttingen}
  \streetaddress{Wilhelmsplatz 1}
  \city{Göttingen}
  \country{Germany}
  \postcode{37073}
}

\renewcommand{\shortauthors}{Pluzhnikov et al.}

\begin{abstract}
This paper presents a specialized methodology for digitizing and segmenting mathematical documents from zbMATH Open, a comprehensive database of mathematical literature, to enhance machine processing capabilities. Currently, approximately 831,000 documents exist only in scanned volumes, which makes them not machine-processable. Furthermore, these scans often span multiple pages or share pages with other documents and incorporate diverse typesetting techniques, posing challenges for automated processing.

To address these issues, we evaluate various Optical Character Recognition (OCR) tools and document separation techniques, proposing an optimized pipeline that outperforms existing approaches. Our study identifies Mathpix as the most effective OCR tool for LaTeX conversion, demonstrating superior performance based on BLEU and Edit Distance metrics. For document separation, we fine-tune generative Large Language Models (LLMs) and integrate them into a Majority Voting framework, achieving 97.5\% accuracy  when providing the text of the document. Additionally, our method identifies the start and end indexes for 90.6\% of the test dataset, with an accuracy of 98.4\% on applicable cases, resulting in an overall accuracy of 89.1\% on the entire dataset. This approach surpasses traditional baselines, including regular expressions, ChatGPT-4o, and computer vision-based techniques.
As a practical outcome, we process 810,977 mathematical documents into machine-readable text and extract precise document boundaries for 721,288 documents in LaTeX format. These contributions significantly improve accessibility for mathematical information retrieval systems, machine learning models, and related applications.
\end{abstract}

\maketitle

\section{Introduction}

Printed mathematical literature has a long history, with records dating back several centuries. Among the most comprehensive resources in the field, zbMATH Open (1868 – present) serves as a critical abstracting and reviewing service in pure and applied mathematics \cite{Mihaljevic2020}. It offers a vast collection of abstracts, reviews, and summaries of mathematical research articles, making it an indispensable tool for researchers in mathematics and related disciplines. As of April 2024, zbMATH Open includes over 4.8 million entries, consisting of abstracts, reviews, and summaries of mathematical research papers. Throughout this paper, we refer to these entries (without metadata) as documents.

\begin{figure}[htbp]
    \centering
    \begin{subfigure}[b]{0.3\textwidth}
        \includegraphics[width=\textwidth]{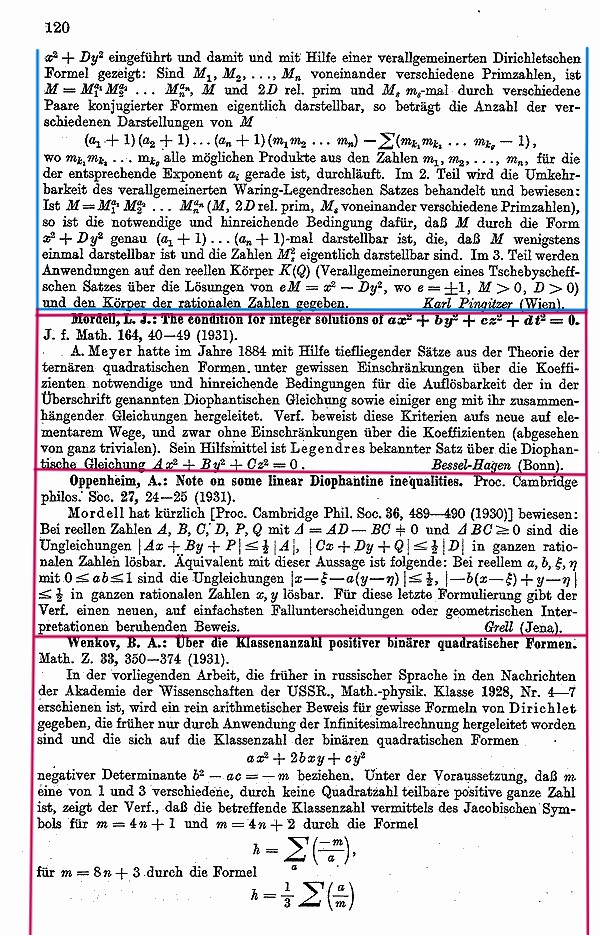}  % Ensure correct path and filename
        \caption{First scanned page}
        \label{fig:subfig1}
    \end{subfigure}
    \hfill
    \begin{subfigure}[b]{0.3\textwidth}
        \includegraphics[width=\textwidth]{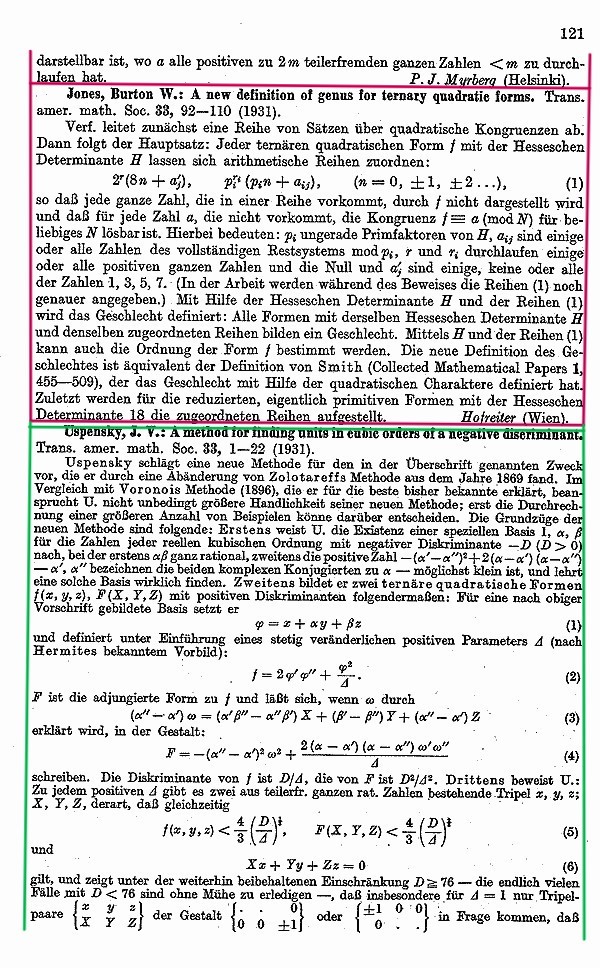}  % Ensure correct path and filename
        \caption{Second scanned page}
        \label{fig:subfig2}
    \end{subfigure}
    \caption{Illustration of two scanned pages from zbMATH Open, showcasing the end of one document (blue), four complete documents (red), and the start of another (green). \href{https://zbmath.org/?q=an\%3A0001.12001}{https://zbmath.org/?q=an\%3A0001.12001}.}
    \label{fig:images_ex_of_scanned}
\end{figure}

However, approximately 831,000 documents in zbMATH Open exist only in scanned volumes (e.g., Figure ~\ref{fig:images_ex_of_scanned}), some of which date back to the 19th century. These scanned documents were digitized from typeset records \cite{Beck2020} and include essential metadata such as titles in LaTeX, publication source, year of publication, and page numbers. Although the scanned volumes allow mathematicians to read individual documents, they are not searchable or suitable for text mining. Mathematical content, such as LaTeX representations of text, formulas, and symbols, is essential to enabling machine processing of mathematical concepts. This, in turn, benefits applications like neural networks and Math Information Retrieval (MathIR) systems \cite{GreinerPetter2020, Kamali2010, Kamali2013, Davila2017}, which are critical to improving the efficiency and accuracy of searches for related mathematical concepts and to assisting in various research processes. For example, MathIR systems allow for "linking \( P_n^{(\alpha, \beta)}(x) \) with ‘Jacobi polynomial’" \cite{Davila2017}. Beyond entity linking, these systems support a range of applications, including plagiarism detection, search engines, and recommendation systems. Therefore, converting scanned images to machine-readable formats is crucial to enhance accessibility and enable advanced data processing.

Beck et al. \cite{Beck2020} conducted an initial study on converting scanned zbMATH Open documents to LaTeX and splitting them for further use. They identified two major challenges: first, scanned pages often feature text in varying styles, heights, and line spacings, and typesetting techniques change over time \cite{schubotz2019four}. Any solution to handle these data must be adaptable to these variations. Second, documents may span multiple scanned pages or multiple documents may appear on a single page (e.g., Figure ~\ref{fig:images_ex_of_scanned}). These challenges highlight the need for specialized methods to accurately split documents. Several studies \cite{Tolokin2021, Tang2024, Zhang2023} have explored document data extraction, employing techniques such as ChatGPT and regular expressions. Recent advancements in Large Language Models (LLMs) offer promising results in document splitting tasks. However, these approaches have not yet been applied to the task of splitting zbMATH Open documents. This work aims to address the challenge of accurately extracting LaTeX documents from scanned zbMATH Open volumes.

The goal of this research is to split the Optical Character Recognition (OCR) output (LaTeX representation of scanned documents) from the zbMATH Open volumes and associate the resulting output with the metadata of the corresponding documents in the zbMATH Open database.

This paper makes the following key contributions.
It introduces a problem-specific methodology for automatic document splitting by leveraging modern generative LLMs. These models, which have shown significant success in natural language processing tasks, are adapted for mathematical text separation, achieving high accuracy in data processing.
The research evaluates OCR tools and document splitting techniques using established metrics such as BLEU and Edit Distance for OCR engines, and accuracy for document splitting. These metrics offer a quantitative assessment of the proposed methods' performance.
Comparative analysis demonstrates that the proposed approach outperforms existing methods, including ChatGPT-4o, computer vision-based approaches, and regular expression-based methods, in terms of key performance indicators, thus highlighting its potential for broader applicability in the field.
As a practical outcome, the methodology successfully processes and converts 810,977 mathematical documents (as text) and 721,288 documents (as LaTeX representations with start and end indexes) into machine-readable and searchable formats.

\section{Related Work}
Several studies have explored the use of regular expressions to extract content from OCR-processed documents. For instance, Pitou et al. \cite{Pitou2016} utilize regular expressions to categorize elements within scanned invoices processed via OCR, identifying details such as account numbers, dates, and amounts. They report a recall of 35.88\% and a precision of 50.46\%. Similarly, Pandey et al. \cite{Pandey2023} employ regular expressions to extract content from a variety of document types after applying OCR. They achieve accuracies of 97\%, 71.87\%, and 78.95\% for different document types.

In other studies, computer vision (CV)-based methods focus on identifying key elements within an image, such as "Text" or "Title", for further use with OCR tools. Li et al. \cite{Li2022} apply the Faster-RCNN model \cite{ren2016faster} to detect the locations of cells within tables. Patil et al. \cite{Patil2022} employ the U-Net architecture \cite{ronneberger2015u} to identify areas containing relevant text before applying the OCR engine. In the application phase, the F1 score achieved is reported as 99.9\%. Furthermore, Zhu et al. \cite{zhu2022docbed} test several models for layout segmentation of newspaper articles. The best performance in terms of the mean Intersection over Union (IoU) metric, which is 70.02, is achieved using the Segmentation Transformer model \cite{zheng2021rethinking}.
According to the Hugging Face documentation\footnotemark[1], the state-of-the-art model for document layout analysis is the Document Image Transformer (DiT) \cite{li2022dit}. This model can classify document elements such as "Text" and "Title". By leveraging advanced architectures like Mask R-CNN \cite{he2017mask} and Cascade R-CNN \cite{cai2018cascade}, and trained on the comprehensive PubLayNet dataset \cite{zhong2019publaynet}, DiT processes document images with multiple levels of detail. In our study, we utilize DiT to compare its performance with our proposed method.

\footnotetext[1]{\href{https://huggingface.co/docs/transformers/model_doc/dit}{https://huggingface.co/docs/transformers/model\_doc/dit}
}

Neural network techniques, designed for natural language processing (NLP), have also been applied to extract text from OCR-processed documents. Palm et al. \cite{Palm2017} describe the use of the LSTM model \cite{hochreiter1997long} on sequential N-grams of text derived from OCR-processed invoices, achieving an average F1 score of 89\% on seen invoice templates and 84\% on unseen templates. In another study, Kruse et al. \cite{Kruse2021} combine OCR with the BERT model to extract definitions from scanned documents, classifying sentences with an F1 score of 79\%. Zhiwei et al. \cite{ma2024information} use GPT-3.5 to extract content from historical documents with different structures after OCR processing, reporting an accuracy of 100\%. Similarly, Perot et al. \cite{perot2023lmdx} use the same model for information extraction from OCR-processed documents, obtaining a Micro-F1 score of 63.86\% on a mixed template of registration forms.

These studies highlight the potential of using generative LLMs for document splitting and content extraction. The successful application of LLMs in related tasks provides a strong baseline for our work, particularly for the task of splitting documents. The use of non-generative methods, such as assigning labels to input sequence elements \cite{Barrow2020, Kruse2021}, is an alternative approach but requires significantly more manual effort, as all tokens in the input text must be marked. In contrast, generative models offer the advantage of correcting OCR errors and removing extraneous meta-information (e.g., page numbers), making the results more user-friendly.

\section{Datasets and metrics}

In this study, we utilized a dataset derived from scanned volumes of mathematical literature. The dataset collection and preprocessing process is detailed as follows:

\subsection{zbMATH Open dataset}

The primary dataset consists of 478 distinct scanned books, encompassing texts in English, German, French, and Italian. The books are divided into pages, each in a separate file, in total 254,444 pages. We also used the following information about 813,005 documents from these books (e.g., Table \ref{table:math_publication}):

\begin{itemize}
    \item \textbf{Scan Document ID}: A unique identifier linking each entry to a specific page.
    \item \textbf{zbMATH Internal ID}: A unique identifier for each document.
    \item \textbf{Title}: The title of the document, which may be translated.
    \item \textbf{Original Title}: Recorded if the Title is a translation.
    \item \textbf{Source}: The venue of publication.
\end{itemize}

\begin{table}[h!]
\caption{Sample of the data from the zbMATH Open.}
\centering
{\fontsize{4.5}{4.5}\selectfont
\begin{tabular}{|c|c|c|c|c|}
\hline
\textbf{Scan\_document\_id} & \textbf{zbMATH\_internal\_id} & \textbf{Title} & \textbf{Original\_title} & \textbf{Source} \\
\hline
001/005 & 3000000 & \makecell{Methods of \\ mathematical \\ physics. \\ Vol. 1. \\ 2nd ext. ed.} & \makecell{Methoden der \\ mathematischen \\ Physik. Bd. I. \\ 2. verb. Aufl} & \makecell{Die Grundlehren der \\ mathematischen \\ Wissenschaften in \\ Einzeldarstellungen \\ mit besonderer \\ Berücksichtigung der \\ Anwendungsgeb. \\ 12. Berlin: Julius \\ Springer. XIV, 469 \\ S. u. 26 Abb. \\ (1931).} \\
\hline
001/005 & 3000001 & \makecell{On the \\ construction of \\ arithmetics} & \makecell{Über den \\ Aufbau der \\ Arithmetik} & \makecell{Jahresber. Dtsch. \\ Math.-Ver. 40, \\ 22-37 (1931).} \\
\hline
001/006 & 3000002 & \makecell{Remarks on a \\ determinant \\ theorem \\ by Minkowski} & \makecell{Bemerkungen zu \\ einem \\ Determinantensatz \\ von Minkowski} & \makecell{Jahresber. Dtsch. \\ Math.-Ver. 40, \\ 49-53 (1931).} \\
\hline
\end{tabular}
\label{table:math_publication}}
\end{table}

To evaluate different document splitting approaches, we prepared a train dataset of 1000 random manually collected samples as well as a test dataset of 478 random documents (one per volume).

\subsection{OCR evaluation dataset}

An evaluation of the different OCR tools requires a suitable dataset that accurately reflects the challenges presented by the mathematical literature. Also, it should consist of "Page" - "LaTeX representation" (as ground truth) pairs. 
To the best of our knowledge (as of 07.2024), there is no dataset already collected that meets the specific needs of this research \cite{Blecher2023}.
Consequently, we collected our own dataset comprising 25 pages sourced from a variety of STEM articles within the Arxiv dataset \cite{Clement2019}, which offers a rich collection of scientific papers in LaTeX format. Of these pages, 22 contain mathematical content, such as formulas. The LaTeX code associated with each page was manually extracted, ensuring a high standard of accuracy to serve as the ground truth in our evaluations.

\subsection{Metrics}

In this paper, when evaluating existing OCR systems that process LaTeX outputs, we applied the BLEU metric \cite{Wei2024, Sun2024, Wang2020, Wang2019, Wang2021, Blecher2023}, which focuses on n-gram precision, and Edit Distance \cite{Wei2024, Sun2024, Wang2020, Blecher2023}, which measures textual similarity between generated and reference outputs.

In the articles \cite{Tang2024, ma2024information} authors, like us, employ generative LLMs to solve the documents splitting task, with accuracy serving as the main metric to evaluate quality. In this case, accuracy is the proportion of correctly processed documents among all documents for which processing was performed. For the document splitting task, we applied this metric.

To properly use the accuracy metric, we must define what is considered a correctly processed document (ground truth) in our splitting task for different types of output. In case the text of the document is the output, the model should provide all LaTeX code from the OCR output located between the first and last characters of the document text, and only it. Among the modifications, OCR errors can be corrected (for example, removing extra spaces before punctuation marks), and metadata can be removed (for example, the page number) from the text. If the document start and end indexes are output, they should point respectively to the first and last character of the document text in the LaTeX representation. If there is no document, "not reviewed." should be returned.

\section{Document Splitting}

\subsection{OCR comparison}

In this section, we provide the evaluation results for the OCR tools (Table \ref{table:ocr_eval}).
The results indicated that Mathpix outperforms InftyReader, Marker, and Nougat in BLEU and Edit Distance. Marker shows the second-best result in terms of BLEU and Nougat in terms of Edit Distance. InftyReader consistently lagged behind the other three tools. 

\begin{table}[h!]
\centering
\caption{Performance metrics of the OCR systems.}
{\fontsize{6}{6}\selectfont
\begin{tabular}{|>{\raggedright\arraybackslash}c|>{\centering\arraybackslash}p{1.2cm}|>{\centering\arraybackslash}p{1.2cm}|>
{\centering\arraybackslash}p{1.2cm}|>{\centering\arraybackslash}p{1.2cm}|}
\hline
\textbf{Metric} & \textbf{Mathpix} & \textbf{InftyReader} & \textbf{Nougat} & \textbf{Marker} \\
\hline
Average BLEU & \cellcolor{green!25}0.7237 & 0.5422 & 0.5666 & 0.5723 \\
\hline
Average Edit Distance & \cellcolor{green!25}0.1733 & 0.2219 & 0.1912 & 0.2046\\
\hline
\end{tabular}}
\label{table:ocr_eval}

\end{table}

Based on the evaluations performed, Mathpix was selected to convert the scanned images to LaTeX. Better scores than other tools on all metrics underscored its ability to better handle this task. The choice of Mathpix provided the highest quality of processed documents available to us.

\subsection{Conversion and challenges}

To facilitate the conversion process, we utilized the Mathpix API, an interface that allows the submission of images and the retrieval of the corresponding LaTeX code. Our dataset comprised 254,444 images in .tiff format, representing scanned pages from zbMATH Open volumes.

Approximately 0.88\% (2,235 images) of the total submissions resulted in API errors. These errors were primarily due to the inability of the OCR system to recognize image content (1,408 cases). Upon errors analysis, we found that we submitted blank sheets of paper as input. In the remaining 827 cases (approximately 0.33\% of all images), Mathpix failed to process the submitted scans. Thus, we received the data necessary for further work, saving each converted scan into a separate .json file - 252,209 in total.

\subsection{OCR errors}
\label{sec:OCR errors}

OCR systems, while advanced, are not perfect. Converting scanned documents with mathematical content into LaTeX format can lead to various errors. Understanding them was important for developing the next steps of this research. This section describes common types of errors.

\begin{itemize}

\item OCR systems may incorrectly interpret characters, especially those that are visually similar. For example, '0' (zero) may be recognized as 'O' (capital letter O), or umlauted characters (e.g., 'ä', 'ö', 'ü') may be confused with their non-umlauted counterparts ('a', 'o', 'u'), or " can be mistaken for 66.

\item OCR output may contain inconsistent line breaks and spacing. For example, the end of a line in the scanned image might not appear in the LaTeX code. There may also be cases of double space in the LaTeX code.

\item The OCR output may include unnecessary LaTeX tags and symbols. As an example, writing "A title of the article" as "\textasciitilde A \textasciitilde t i t l e \textasciitilde o f \textasciitilde t h e \textasciitilde a r t i c l e \textasciitilde".

\item The OCR system may skip some characters and entire words.

\end{itemize}

The presence of such errors, which are quite common, leads to the complication of further work. Especially when it comes to approaches based on regular expressions.

\section{Regular expressions approach}

Several approaches based on the regular expressions (regex) have demonstrated success in extracting data from outputs of the OCR systems. Therefore, we consider regex as one of the baselines for document splitting. This section outlines the methodology, challenges, and results of applying regular expressions in this context.

\subsection{Methodology}
\label{sec:OCR_method}
The objective was to automate the extraction of individual documents from OCR-processed files in LaTeX.

Steps involved:

\begin{enumerate}

\item {For each $row$ in the zbMATH Open table
\ref{table:math_publication}:}

\item {\bf Reading next row:}
Read the next row from the zbMATH Open table as the $next\_row$.

\item {\bf Reading LaTeX files:}
Read the $LaTeX\ content$ of the files from $row$[Scan document id] to $next\_row$[Scan document id] inclusive. If there is more than one file, connect the LaTeX code sequentially. 

\item {\bf Search for the original title and title:} Search for the $row$[Original title] in the $LaTeX\ content$. If there is no match, search for $row$[title]. We denote the index of the end of the match as $start$.

\item {\bf Next\_row title and original title search:}

Repeat the previous step for $next\_row$.  We denote the index of the beginning of the match as $end$.

\item {\bf Text extraction part:} Extract the text between $start$ and $end$ from $LaTeX\ content$. We denote the extracted text as $text$.

\item {\bf Removing extraneous text:} Remove extra parts from $text$.

\item {\bf Saving extracted text:}Save the $text$ as\\[0.0em]
\hspace*{0.00em}$row$[Abstract/Review/Summary].

\end{enumerate}

Steps 4 and 5 were necessary to find the tentative boundaries of the searched document. In Step 7, these bounds were refined using regular expressions. This helped to strip away superfluous elements such as page numbers, author names, parts of the title, and other metadata, leaving only the document's text. The applied regular expressions were iteratively changed to handle the next detected algorithm or OCR system error. In the case when the document was the last in the book, we artificially added another title to the end of the text and applied the algorithm.

\begin{figure}
    \centering
    \includegraphics[width=0.4\textwidth]{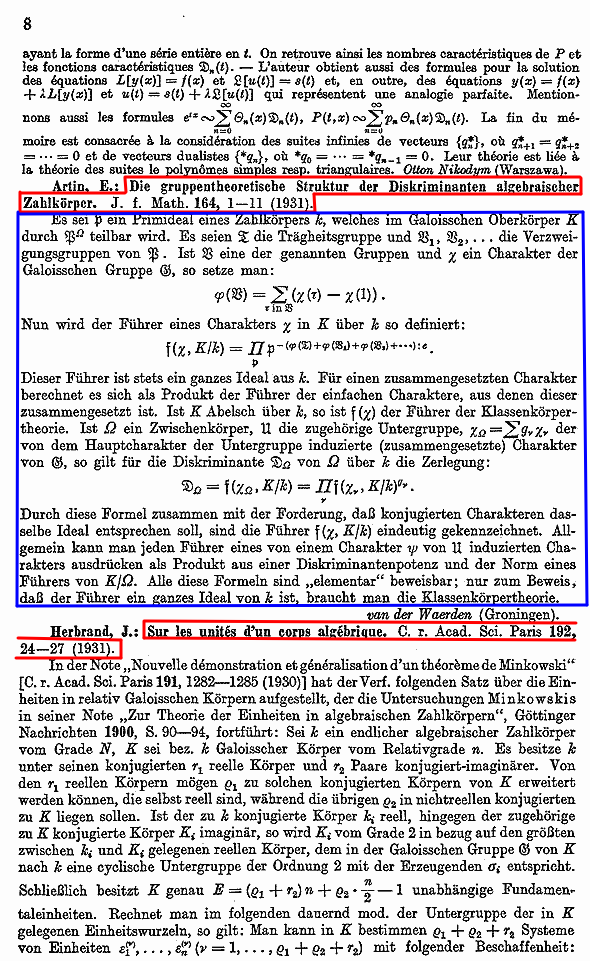}
    \caption{RE-based extraction approach. Red - titles and sources matches. Blue - text to be extracted.}
    \label{fig:re_example}
\end{figure}

An example can be seen in Figure ~\ref{fig:re_example}. 

\subsection{Exclusion of documents from consideration}
\label{sec:Exclusion of documents}
Taking into account the proposed algorithm, we took the following step to clean the dataset before applying it. We excluded those documents on which it would not have worked due to the absence of certain data.

 If for any page between $row$[Scan document id] and $next\_row$[Scan document id] inclusive LaTeX representation is missing, the document was excluded from further consideration. There were 2,028 such documents in total.

After this, 810,977 of the 813,005 documents remain. For them, we apply the algorithm presented.

\subsection{Challenges}

The application of the described approach presented several challenges due to the OCR errors, the variability in document formatting (for example, the presence or absence of a typed ID before the document) and the possibility of writing the same formula in different ways. Because of this, exact matches between titles from the data table and from the LaTeX text often did not occur during the steps 4 and 5 of the algorithm. In such cases, we used regular expressions in an attempt to eliminate typical OCR errors. After that, steps 4 and 5 of the algorithm were repeated once again. As a result, the algorithm suggested an answer for 693,635 documents (85.5\%).

To increase that number, we allowed the Levenshtein Distance \cite{Levenshtein1966} between the desired title and the section of LaTeX code to be 0.3 times the length of the title, rather than requiring an exact match. This threshold was chosen because, at this value, the growth in the number of documents for which the approach could produce an answer was significantly reduced. We expected that step 7 of the algorithm would reduce the drop in the accuracy metric caused by this action.

\subsection{Results}

The algorithm returned extracted text for 786,314 out of 810,977 entries (97\%). However, the described challenges led to inaccuracies in text extraction, complicating the process of removing extra information except for the document itself. Further improvement (modification of the regular expressions) of the algorithm requires significant manual labor to analyze all errors that occur and write rules to eliminate them.

\section{LLM-voting based document splitting}

 Several approaches based on the generative LLMs have shown success in document splitting tasks. Hence, we used generative LLMs as possible candidates for the solution. 

\subsection{Methodology}\label{sec:LLMs methodology}
The idea of the approach is as follows:

\begin{enumerate}

\item For each $row$ in the zbMATH Open table

\item Read the next row from the zbMATH Open table as the $next\_row$.

\item Add to the prompt instructions for extracting the document, $row$[Original title] if exist (otherwise $row$[Title]), $row$[Source]. 

\item Add to the prompt  $next\_row$[Original title] if exist (otherwise $next\_row$[Title]), $next\_row$[Source].
\item Implement the step 3 from the section~\ref{sec:OCR_method}. Add the result to the prompt.

\item Inference of the generative LLM with the resulting prompt.
\end{enumerate}

\begin{figure}[h]
    \centering
    \includegraphics[width=1\linewidth]{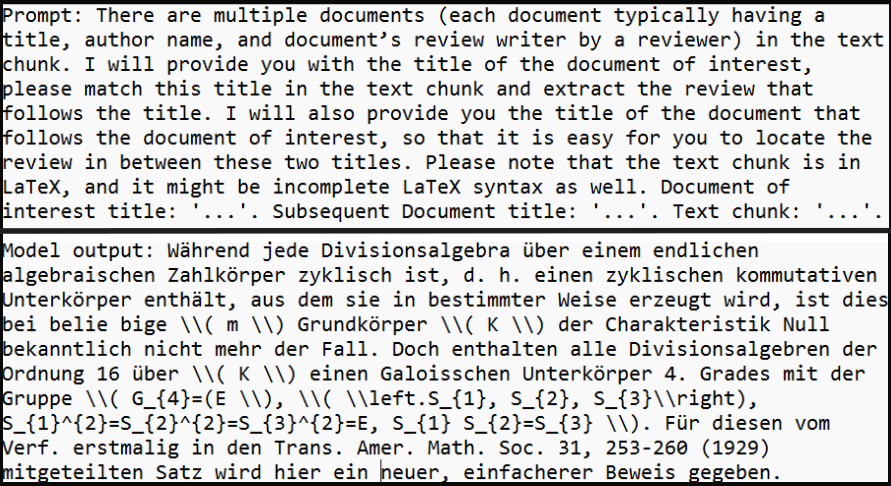}
    \caption{Example of the prompt and correct generative Large Language Model output for the document {\href{https://zbmath.org/0005.34202}{https://zbmath.org/0005.34202}}.}
    \label{fig:prompt}
\end{figure}

In the case when the
document was the last in the book, we followed the procedure outlined in section~\ref{sec:OCR_method}.
 An example of the prompt can be seen in Figure ~\ref{fig:prompt}. We also applied the steps described in section~\ref{sec:Exclusion of documents} to the zbMATH Open table.

\subsection{Base models}

\footnotetext[1]{\href{https://llm.extractum.io/}{https://llm.extractum.io/}}

The next step was to choose a model. Due to the technical limitations of the
GPU, we opted for models that require less than 16GB of VRAM. The size of
the context window was also critical. To avoid the situation where the model
may not be able to handle the text of the document due to its size, it had to be more than 4 thousand tokens (which corresponds to approximately 8
pages of plain text).

To search among open-source generative models, we used the LLM Explorer website\footnotemark[1] (as of 30.04.2024) which provides a convenient interface for
comparing generative LLMs available on Hugging Face. As part of our project, we decided to focus on the Instruction-Based Large Language Models
because it best represents what we want the model to accomplish. In addition, there are scores for the models on the website (Hugging Face average
score). This information was also an important criterion in the selection of
models.

We selected two models with the highest score and one model with the highest score among the models with the largest number of parameters that we could use (as potentially more complex). The characteristics of the selected models are provided in Table ~\ref{table:models} (hereafter, we refer to them as "base" models).

\begin{table}[h!]
\centering
{\fontsize{7}{7}\selectfont
\begin{tabular}{|l|c|c|c|c|}
\hline
\textbf{Name} & \textbf{Score} & \textbf{Context Length} & \textbf{Size} & \textbf{VRAM (GB)} \\
\hline
CalmExperiment-7B-slerp\footnotemark[1]
 & 76.67 & 32k & 7B & 14.4 \\
\hline
Maxine-7B-0401-Stock \cite{MaxineStock} & 76.73 & 32k & 7B & 14.4 \\
\hline
Mistral-Passthrough-8L-10B\footnotemark[2] & 71.01 & 32k & 10B & 14.5 \\
\hline
\end{tabular}}
\caption{Large Language Models overview.}
\label{table:models}
\end{table}

\footnotetext[1]{\href{https://huggingface.co/allknowingroger/CalmExperiment-7B-slerp}{https://huggingface.co/allknowingroger/CalmExperiment-7B-slerp}}
\footnotetext[2]{\href{https://huggingface.co/DeepKarkhanis/Mistral-Passthrough-8L-10B}{https://huggingface.co/DeepKarkhanis/Mistral-Passthrough-8L-10B}}

\subsection{Fine-tuning}

To significantly reduce the resources spent adapting the Large Language Models to specific the task, we used the fine-tuning technique \cite{Krizhevsky2012}. Instead of training neural networks from scratch, we took pre-trained ones. Since the models learned to solve NLP problems, they were already able to work with the basic concepts of the language. In this case, we needed much less time, computing resources, and data to achieve a good result. To speed up fine-tuning of Large Language Models, we used the LoRa (Low-Rank Adaptation) optimization technique \cite{Hu2021}. For implementation of the fine-tuning, we utilized LLaMA-Factory \cite{zheng2024llamafactory}. 

\begin{figure}
    \centering
    \begin{subfigure}[b]{0.5\textwidth}
        \centering
        \includegraphics[width=\textwidth]{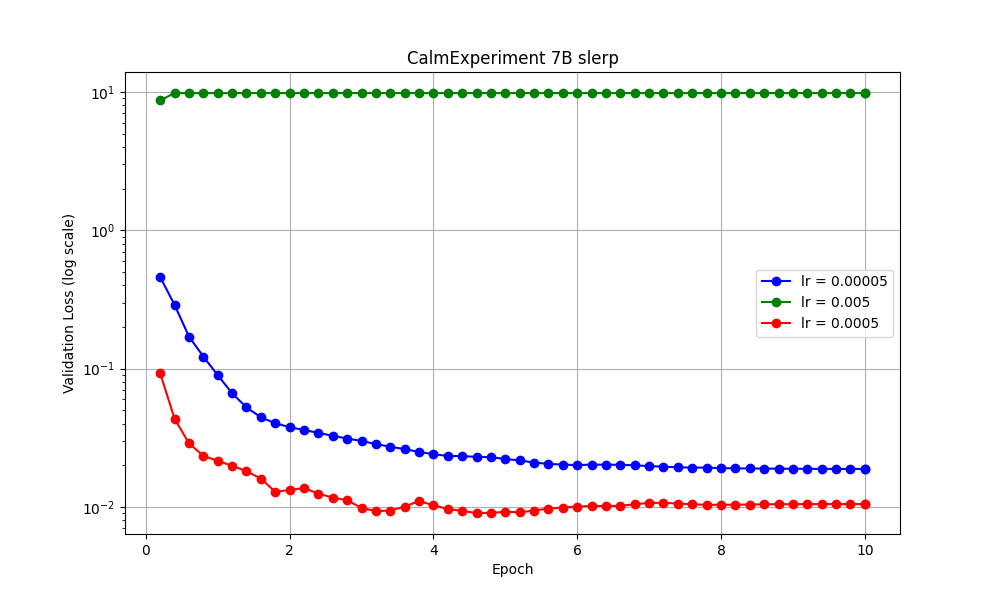}
        \caption{CalmExperiment-7B-slerp validation loss.}
        \label{fig:image1}
    \end{subfigure}
    \vfill
    \begin{subfigure}[b]{0.5\textwidth}
        \centering
        \includegraphics[width=\textwidth]{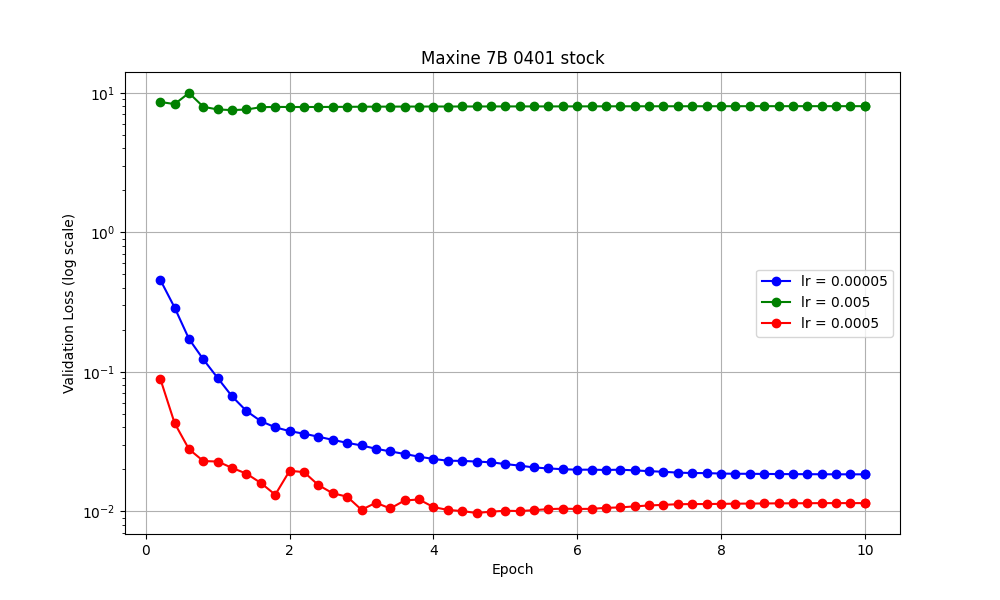}
        \caption{Maxine-7B-0401-stock validation loss.}
        \label{fig:image2}
    \end{subfigure}
    \vfill
    \begin{subfigure}[b]{0.5\textwidth}
        \centering
        \includegraphics[width=\textwidth]{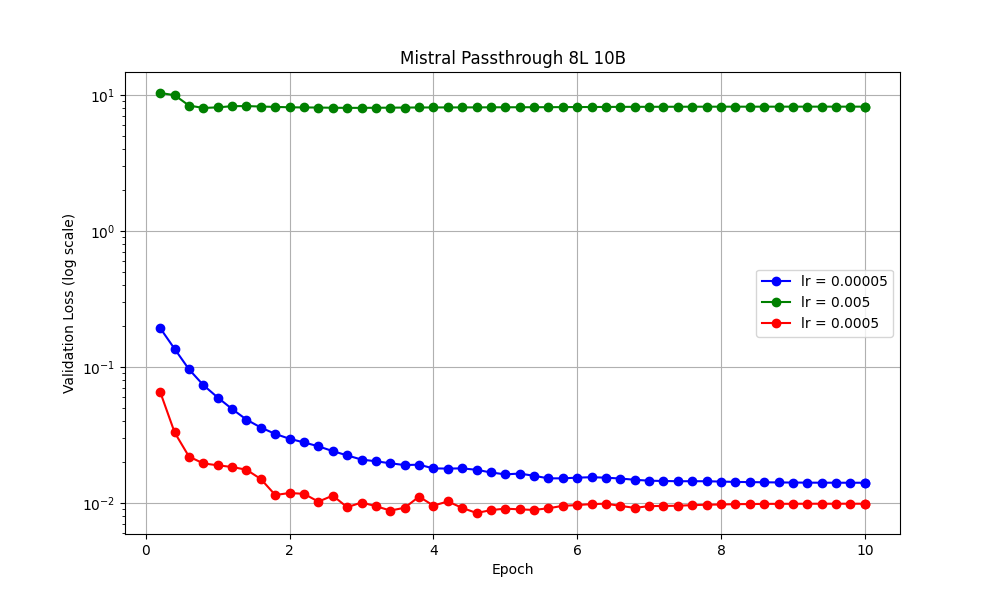}
        \caption{Mistral-Passthrough-8L-10B validation loss.}
        \label{fig:image2}
    \end{subfigure}
    \caption{Validation losses of the Large Language Models.}
    \label{fig:validation}
\end{figure}

To select the optimal hyperparameters, such as the learning rate and the number of epochs for fine-tuning, we validated the models (with the cosine learning rate scheduler and batch size equal to 2 as the largest number we could use due to the GPU limits) by splitting the train dataset in a ratio of 80\% (for training) and 20\% (for validation). The split was made at random. The validation results can be seen in Figure ~\ref{fig:validation}.

Based on the data obtained, the learning rate was chosen to be 0.0005 for all models, since at this rate the losses for all of them are the minimum of those presented. The number of epochs for training all models was determined to be 5 because at this point the loss value stabilized and stopped decreasing. Upon reaching this epoch, CalmExperiment-7B-slerp shows the smallest loss of all models. 

\begin{figure}[h]
    \centering
    \includegraphics[width=1\linewidth]{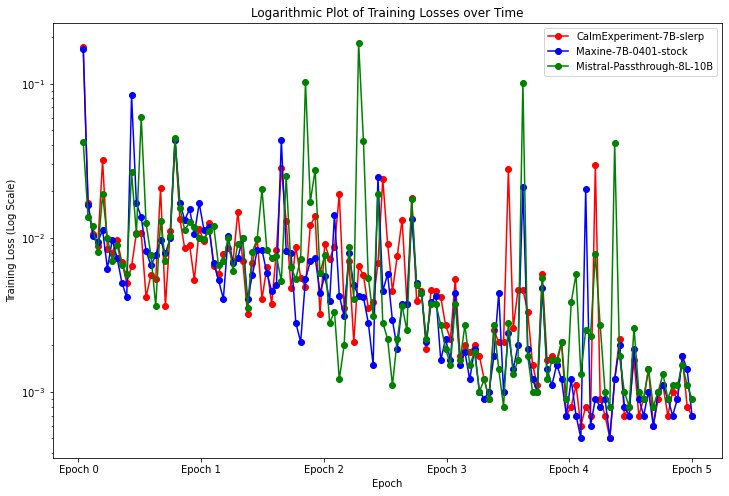}
    \caption{Training losses of the Large Language Models.}
    \label{fig:finetuning_loss}
\end{figure}

Taking into account all the hyperparameters described above, we performed a fine-tuning of all three models. All 1000 samples from the train dataset were used.
The training loss can be seen in Figure ~\ref{fig:finetuning_loss}.

\subsection{Majority Voting approach}

In addition to using each model individually, we have implemented an approach that allows them to be used simultaneously - to reduce the probability of mistakes by accounting for multiple predictions. The popular technique of model assembly was used, Majority Voting \cite{re2012ensemble}. The essence of this method is as follows:

\begin{enumerate}

\item For every document, each of the models proposes its output.

\item If the output of any 2 models matches, this result is considered final.

\item If all 3 outputs are different, we take the output of the model that performed better on the validation dataset - CalmExperiment-7B-slerp. Hereafter, we refer to it as the "best" model.

\end{enumerate}

Regarding just using the "best" model, it can help in the following way. If the "best" model happens to make an error, the collaborative nature of the Majority Voting system allows for errors to be corrected by leveraging consensus from the other models. At the same time, there may be a chance of making the result worse than just using the "best" model, but it requires that the other two models make the same mistake in the document splitting process, and the answer of the "best" model is simultaneously correct. 

\subsection{LLM baseline}

We used the latest version of ChatGPT-4o (as of 08.2024) as a baseline, as it has shown success in similar tasks performed by Zhiwei et al. in the work \cite{ma2024information} and Perot et al. in the work \cite{perot2023lmdx}. To do this, we had to modify the prompt by explicitly specifying the expected behavior in the absence of the document content by adding the text "If the review is empty, return ‘Not reviewed’" (fine-tuned models received this information from the train dataset). In addition, we also applied an approach based on the use of a prompting technique known as few-shot learning (the essence of the technique is to include a small number of labeled examples in the prompt), as it can greatly improve the performance of the model \cite{brown2020language}. In our case, we include two labeled examples.

In addition, we tested the "base" models on a similarly modified prompt (without few-shot learning) before fine-tuning.

\section{Computer vision-based baseline}

Given the success of computer vision models in document splitting tasks, we selected a CV-based approach as a baseline. When using such approaches, an important aspect is the ability of the model to distinguish different parts of the document. More precisely, the model should be able to determine a title from plain text. The proportion of correctly identified titles among all actual titles serves as an upper bound for the accuracy of the entire CV-based document splitting method. This is because titles are used to define the boundaries of the document of interest (as there may be no text relating to the document at all on the given page).

\begin{figure}[h]
    \centering
    \includegraphics[width=0.8\linewidth]{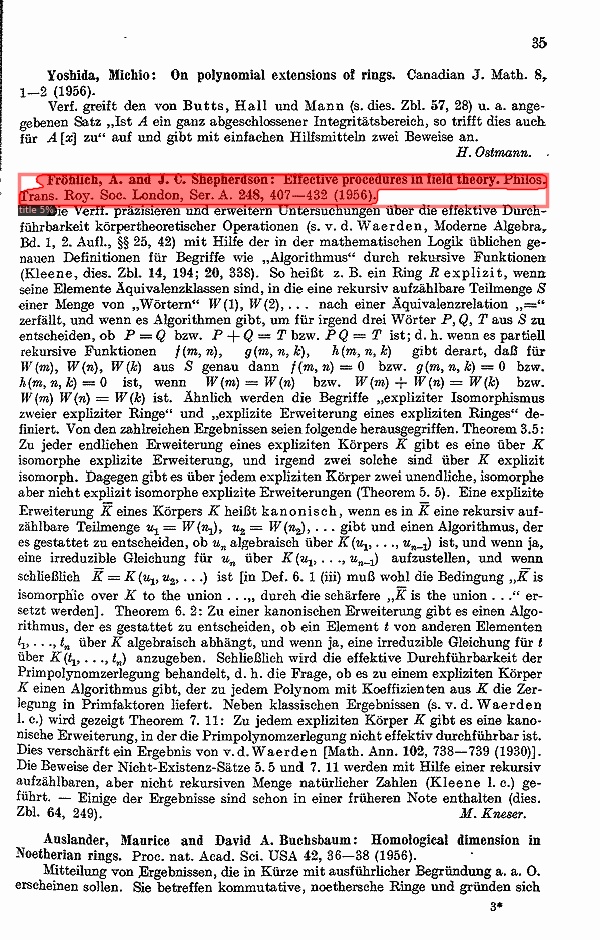}
    \caption{CV model output on a page with 3 titles, of which 1 was recognized.}
    \label{fig:layout_model}
\end{figure}

In this research, we used a state-of-the-art Document Image Transforme (DIT) model  \cite{li2022dit} for document layout analysis, evaluating its performance on the test dataset. To do this, we modified the original code, leaving only the "title" label (originally there was "text", "title", "list", "table", and "figure"). An example of the model output can be seen in Figure ~\ref{fig:layout_model}.

\section{Evaluation and all data processing.}

\subsection{Evaluation of "base" models}

To assess the performance of the "base" models prior to fine-tuning, we evaluated their initial versions on the document splitting task. This allowed us to quantify the improvements achieved through fine-tuning by comparing the resulting accuracy. For this, we used the test dataset.

Our results showed that none of the models returned outputs that met the accuracy criteria. The models returned a lot of unnecessary text (e.g. trying to continue text from scanned pages or keeping the metadata about the document), despite the refinements in the prompt. In addition, the searched documents were often not even partially included in the output. Thus, the accuracy metric for "base" models without fine-tuning is 0\%. We conclude that these models are unsuitable for solving our task without fine-tuning.

\subsection{Evaluation results}

This section presents a comprehensive evaluation on the test dataset of the approaches developed by us, as well as ChatGPT-based and CV-based. The results are shown in Table ~\ref{table:eval_results}

\begin{table}[h!]
\centering
\begin{tabular}{|l|c|c|c|c|}
\hline
\textbf{Approach} & \textbf{Accuracy} \\
\hline
Regular Expressions  & 68.8\%\footnotemark[1]\\
ChatGPT-4o  & 51 \%\\
ChatGPT-4o Few Shot Learning & 52.1 \%\\
DIT  & 48.5\%\footnotemark[2]\\
CalmExperiment-7B-slerp & 96.2\%\\
Maxine-7B-0401-Stock & 93.1\%\\
Mistral-Passthrough-8L-10B  & 93.7\%\\
Majority Voting  & \textbf{97.5}\%\\

\hline
\end{tabular}

\caption{Evaluation results of the document splitting approaches.}
\label{table:eval_results}
\end{table}

Based on the results obtained, we conclude that our proposed approaches leveraging Large Language Models significantly outperform alternative methods, which are based on regular expressions, the widely used ChatGPT-4o model (which often included superfluous metadata in the response), and computer vision techniques. Among the models that we fine-tunde, CalmExperiment-7B-slerp showed the best result, outperforming the others by more than 2\%. However, expectedly the highest accuracy was demonstrated by the Majority Voting approach - 97.5\%.

\footnotetext[1]{No document text was offered for the 12 titles.}
\footnotetext[2]{Accuraccy upper bound.}

\subsection{Errors analysis}

This section reviews the main sources of errors that occur when using the Majority Voting method and analyzes each type. By doing this, we can better understand the limitations of the current approach and guide further optimization efforts to improve the accuracy and reliability of document-splitting methods.

\begin{itemize}

\item \textbf{OCR formatting errors:} an OCR error may cause the document to be processed incorrectly (e.g., including the document author in brackets beginning in the document's text may cause the model to recognize the author as part of the text itself). In the test data, it occurred in 2 cases.

\item \textbf{Ambiguity in similar titles:} in zbMATH Open data, neighboring document titles can be nearly identical, differing only by part numbers or minor variations in wording (e.g., "Invariant Imbedding and Radiation Dosimetry. VI" vs. "Invariant Imbedding and Radiation Dosimetry. V"\footnotemark[1]). This similarity can lead the LLM to mistake one document for the other. In the test data, it occurred in 2 cases.

\item \textbf{Hallucinations in Large Language Models outputs:} Our approach is also susceptible to the problems of any generative LLMs - hallucination \cite{huang2023survey}. The reason is the predictive nature of Transformers. These models are trained to find the most likely sequences of tokens based on huge amounts of text and, in our case, can, at some point, continue the text of the document with tokens that seem logical but are not in the original text. Longer documents are more prone to this problem because they require more tokens to be generated. In the test data, it occurred in 4 cases.

\item \textbf{Metadata issues:} in some cases, the model may mistake metadata for part of the document or vice versa. In such a case, either extra text will be returned or the required text will not be returned. In the test data, it occurred in 4 cases.

\end{itemize}

\footnotetext[1]{\href{https://zbmath.org/0274.92002}{https://zbmath.org/0274.92002}}

\subsection{Error detection hypotheses}

In addition to analyzing specific types of errors, we tested two hypotheses to identify potential indicators of errors in the document splitting process. These hypotheses focus on aspects of OCR quality and the probabilistic nature of Transformers-based Large Language Models, aiming to establish methods for error detection. The fine-tuned CalmExperiment-7B-slerp model, as the best of the "base" models, was used in all experiments below.

\textbf{Hypothesis 1: OCR confidence rate as an indicator of error likelihood.}

The first hypothesis considered the relationship between the accuracy of the OCR and the likelihood of errors in generative LLM outputs. Specifically, we hypothesized that the confidence rate, which is the "estimated confidence of output quality" provided by the OCR system (Mathpix) could be a useful predictor of splitting accuracy. The idea was that OCR quality, as indicated by lower confidence scores, may correlate with worse document splitting outcomes and could signal a higher likelihood of errors. However, our findings indicated that the Mathpix confidence rate did not consistently predict error-prone segments. In practice, error rates did not differ significantly between high- and low-confidence OCR outputs. Of the 18 documents processed with errors, 8 were located on pages with confidence rates above the average on all 478 document pages. At the same time, among the 10 documents located on pages with the lowest confidence rate, only one was processed with an error. This result suggests that OCR confidence alone is not sufficient to reliably forecast issues in the generative LLM-based document splitting method.

\textbf{Hypothesis 2: token probabilities as a predictor of segmentation inaccuracy \cite{varshney2023stitch}.}

The second hypothesis explored the potential of using token probabilities as indicators of segmentation accuracy. Since LLMs operate on probabilistic predictions, each generated token is assigned a probability that reflects the model’s confidence in its choice. We hypothesized that lower token probabilities across a sequence might signal lower confidence and, consequently, a higher likelihood of errors in document segmentation. To test this hypothesis, we modified the code of the LLaMA-Factory tool and collected the probability value of each output token of each output. We calculated the probability of the first token, the average probability of the first five (if exist) tokens, and the average probability of all tokens in the output sequence as possible quality metrics. This method did not work as well. For all 18 documents processed with error, no probability took values other than 1.0 more than 2 times. At the same time, among the 10 documents with the lowest values of each of the probabilities, there were also not even 5 processed with error.

The evaluation of these two hypotheses provided valuable information on the limitations of error detection strategies for the OCR-integrated generative LLM-based document splitting process. Although initial assumptions suggested that OCR confidence scores and token probability metrics might serve as reliable indicators of errors of the suggested approach, our findings did not fully support either as standalone predictors.

\subsection{Obtaining start and end indexes of documents.}

To obtain the start and end indexes of documents, we searched for an exact match of the document text, as provided by our best approach (Majority Voting), within the $LaTeX\ content$. If a match was found, we recorded the start and end indexes of the occurrence. If Majority Voting returned "not reviewed.", we noted this response. If neither of the above conditions was met, we considered the attempt unsuccessful.

To increase the success rate of this approach, we removed spaces from both texts we were working with (later restoring the original start and end indexes of the match). This was done because the OCR system was making errors by inserting extra spaces in the text. LLMs correct these errors, but this affected exact matches.

Applying this approach on the test dataset, we obtained indexes or "not reviewed." for 433 documents - 90.6\% of the whole test dataset. The accuracy metric among them reached a value of 98.4\%. Thus, the overall accuracy on the entire test dataset was 89.1\%. This method did not propose exact matches in the source text for all the texts of the documents whose processing produced hallucinations. Thus, although it works on fewer documents, it achieves high accuracy on those it does work on and filters out cases of hallucinations. In cases where it is preferable to have fewer documents, but with higher accuracy on the available ones, this becomes a key factor. Also, it allows one to get the text of the document from the OCR output without any modifications. Most of the unsuccessful attempts happened because of metadata deletion by LLMs.

\subsection{All data processing}

After achieving a high accuracy score on the test dataset, the next step was to use the best approach we had created on the entire dataset. Thus, we applied the Majority Voting approach to 810,977 elements in the zbMATH Open table, finishing the transformation of these mathematical documents into machine-readable and searchable formats as a practical outcome of this research. The result of Majority Voting differed from the CalmExperiment-7B-slerp model in 3.4\% of the cases.
As a result, we extracted texts for the 810,977 documents. After that, we obtained the start and end indexes of the documents using these texts. This process yielded results for 721,288 documents - 88.9\% of the total number of texts from the previous step.

\section{Conclusion}

This project successfully identified the optimal OCR tool, Mathpix, for converting scanned pages to LaTeX format, using the BLEU Score and Edit Distance metrics. As the next step, 252,624 scanned pages from zbMATH Open were converted into LaTeX. Building on this, we developed a problem-specific method for automatic document splitting using Large Language Models. By employing a Majority Voting mechanism across three fine-tuned generative LLMs, the proposed approach achieved an accuracy rate of 97.5\% on the test dataset when providing the text of the documents, significantly surpassing the baseline methods based on regular expressions, Computer Vision, and ChatGPT-4o. Another advantage of our approach is its ability to correct errors made by OCR, as well as to remove metadata (e.g., page numbers) from document texts. We also identified the start and end indexes of the documents for 90.6\% of the test dataset. On the subset where the method was applicable, we achieved accuracy of 98.4\% with 89.1\% accuracy on the whole test dataset.

This work contributes several advancements:

\begin{enumerate}
    \item A pipeline for digitizing and splitting documents with mathematical content from scans.

    \item Demonstrating generative LLM's ability and superiority in handling complex document splitting tasks.

    \item As a practical outcome, we made 810,977 zbMATH Open documents (as text) and 721,288 (as start and end indexes in LaTeX representation of scans) machine-readable and searchable. These documents are now available for applications such as Math Information Retrieval systems, search engines,  plagiarism detection, and neural network training.
\end{enumerate}

The results of the research, including code, models, and datasets, have been made publicly available\textsuperscript{1, 2, 3, 4, 5, 6}, encouraging future research and practical implementations in similar domains.

\footnotetext[1]{\href{https://github.com/zbMATHOpen/retroD/tree/master/documents_splitting/Thesis_Pluzhnikov_Ivan}{https://github.com/zbMATHOpen/retroD/tree/master/documents\_splitting/Thesis\_Pluzhnikov\_Ivan}}

\footnotetext[2]{DOI: 10.5281/zenodo.14253145}

\footnotetext[3]{DOI: 10.5281/zenodo.14496349}

\footnotetext[4]{\href{https://huggingface.co/Ivan444410/Mistral-Passthrough-8L-10B-zbmath}{https://huggingface.co/Ivan444410/Mistral-Passthrough-8L-10B-zbmath}}

\footnotetext[5]{\href{https://huggingface.co/Ivan444410/CalmExperiment-7B-slerp-zbmath}{https://huggingface.co/Ivan444410/CalmExperiment-7B-slerp-zbmath
}}

\footnotetext[6]{\href{https://huggingface.co/Ivan444410/Maxine-7B-0401-stock-zbmath}{https://huggingface.co/Ivan444410/Maxine-7B-0401-stock-zbmath}}

\section*{Acknowledgments}
Funded by the Deutsche Forschungsgemeinschaft (DFG, German Research Foundation) –  \href{https://gepris.dfg.de/gepris/projekt/437179652}{437179652}; \href{https://gepris.dfg.de/gepris/projekt/567156310}{567156310}.

\bibliographystyle{ACM-Reference-Format}
\bibliography{sample-base}

\end{document}